\documentstyle[12pt]{article}
\begin{document}
%\draft

% below commands are for enumerating eqs according to sections
%\newcounter{eq}[section]
%\newcommand{\set}{\stepcounter{eq}
%\renewcommand{\theequation}{\mbox{\arabic{section}.\arabic{eq}}}}

% command for enumerating Refs with 1. instead of [1]:
\makeatletter
\renewcommand{\@biblabel}[1]{\quad#1.}
\makeatother

\hsize=6.15in
\vsize=8.2in
\hoffset=-0.42in
\voffset=-0.3435in

\normalbaselineskip=24pt\normalbaselines

\begin{center}
{\large \bf Deciphering neural circuits for {\it C. elegans} behavior by computations
and perturbations to genome and connectome }
\end{center}

\vspace{0.15cm}

\begin{center}
{Jan Karbowski$^{1,2}$}
\end{center} 

\vspace{0.05cm}

\begin{center}

{\it $^{1}$ Nalecz Institute of Biocybernetics and Biomedical Engineering, \\
Polish Academy of Sciences, 02-109 Warsaw, Poland;  \\
 $^{2}$  Institute of Applied Mathematics and Mechanics,  \\
University of Warsaw, ul. Banacha 2, 02-097 Warsaw, Poland.  }
 
% and 
% Division of Biology 216-76, \\
% California Institute of Technology,
% Pasadena, CA 91125, USA }
\end{center}

%\date{\today}

\vspace{0.1cm}

%\widetext
\begin{abstract}
{\it Caenorhabditis elegans} nematode worms are the only animals with the known 
detailed neural connectivity diagram, well characterized genomics, and relatively 
simple quantifiable behavioral output. With this in mind, many researchers view 
this animal as the best candidate for a systems biology approach, where one can
integrate molecular and cellular knowledge to gain global understanding of worm's 
behavior. This work reviews some research in this direction, emphasizing computational 
perspective, and points out some successes and challenges to meet this lofty goal.

\end{abstract}

%\pacs{PACS Nos. 87.18.Sn, 87.19.La, 87.80.Xa, 89.40.+k, 87.23.Ge}
%\narrowtext 
%\maketitle 

%\begin{narrowtext}

%\maketitle


\vspace{0.1cm}

\noindent Email:  jkarbowski@mimuw.edu.pl

%\noindent Phone: (626)-395-5840, Fax: (626)-795-2397.

\vspace{2.3cm}

%\newpage

%\noindent {\bf Author Summary:}

\newpage

\noindent
{\bf Introduction.} \\
Despite diverse goals and scopes, engineering and biological sciences share a common 
general methodology. If engineers want to know how a device works, they usually 
break or decomposes it into smaller parts and study them in isolation \cite{csete}. 
Subsequently, they gradually put the components together and investigate their interactions,
which generally leads to deciphering the workings of the device. A similar version of 
such a reverse engineering is used also in biology. The functions of biological circuits 
are usually decoded by application of either molecular or cellular perturbations in the 
form of genetic mutations or cell elimination (laser ablations), with simultaneous 
observation of their consequences on the system performance \cite{kitano}.

{\it Caenorhabditis elegans} worms are unique biological organisms to study the 
structure-function relationship across different scales, from molecules to
behavior, for the three main reasons. (i) Their genome and protein networks are well 
characterized \cite{bargmann1998,hobert}, which represents a microscopic level.
(ii) They have a very small nervous system composed of only 302 neurons and they 
are the only animals on the Earth with the known wiring diagram of neural connections
\cite{white,chen,varshney}, called connectome, and this represents a mesoscopic level  
(for a comparison human brain contain $10^{11}$ neurons and even a small fruit fly 
brain has $\sim 10^{5}$ neurons). (iii) These worms can exhibit a broad range of 
behaviors (locomotion, olfaction, complex mating, sleep, learning and memory) that can 
be quantified, and this represents a macroscopic level \cite{debono}. It should be
noted that the structure-function relationship has also been successfully studied
in rodents \cite{helmstaedter} and in the fly \cite{takemura}. However, these systems, 
despite recent progress in their connectome studies, still lack a detailed 
neuron-to-neuron wiring diagram.

The above suggests that an integrated description of the nematode worms across different 
spatial scales is in principle possible by merging the tools of separate disciplines 
such as genomics, connectomics, behavioral neurophysiology, and computational biology. 
There is an expectation that such a description might provided insights not only 
about inner mechanisms employed by the worm to execute its behavioral program 
\cite{hobert,debono}, but also can shed some light on
the biological mechanisms in more complex animals because many molecular processes
and their modularity are preserved across different species \cite{kitano,hartwell}.
However, despite the apparent structural-behavioral simplicity of {\it C. elegans} 
nematodes, the system-level approach is not as straightforward as it might seem. 
Foremost it requires a collaboration of researchers with different backgrounds and 
skills who have to learn the basics of the other disciplines to communicate efficiently.
Despite this practical difficulty there are some studies that successfully merge and 
apply the tools from molecular genetics, behavioral neurophysiology \cite{piggott}, 
and/or computational biology \cite{karbowski2008,brown,kato,rakowski2017}. 
On the other hand, such a merging often requires a lot of guess-work, since each of 
the levels (i-iii), although well characterized, contains some ``knowledge gaps'', 
and moreover the levels are not easily related to one another.

The goal of this review is to give a sense and examples of how the above three levels
could be potentially integrated using a system-level computational perspective, with 
hopes and challenges that have to be addressed and solved before systems biology of 
{\it C. elegans} worms can fully materialized. We focus mainly on locomotion as it 
accompanies a large part of the behavioral output of these worms \cite{chalfie,zhen,haspel}. 
Specifically, we consider the questions of locomotion robustness, locomotion encoding 
in {\it C. elegans} nervous system, locomotory decision making, the type of synaptic 
signaling between premotor command neurons, and the question about which neuronal models 
could be best used for behavioral description?

%\newpage
\vspace{0.35cm}

\noindent
{\bf How robust is {\it C. elegans} locomotory behavior?} \\
Sinusoidal locomotion is the basic short-term behavior of {\it C. elegans} nematodes, 
and therefore one can suppose that it must have been somehow optimized during evolution 
(see e.g. \cite{parker,alexander}). Interestingly, most genetic mutations do not have 
a visible effect on locomotion \cite{park}, and similarly, random elimination of
neurons from the worm's network rarely alters the motion. To affect the locomotor output, 
e.g., to change mean velocity, frequency of body undulations, or the rate of changing 
direction (reversals), one has to
apply targeted mutations or neural ablations \cite{karbowski2006}. But even in these 
cases, the proportional relationship between worm's velocity and the frequency of 
neural oscillation is mostly preserved (Fig. 1). The exceptions that break this 
proportionality are rare, but if they happen they can have a very dramatic influence
on motion (i.e. phenotype), including its cessation. The reason for a high degree
of motor robustness against genetic mutations and neuron eliminations is that both
protein and neural networks possess the so-called ``rich club'' architecture
\cite{towlson}. This means that connectivity in these two types of networks is
generally sparse, with only a small fraction of proteins and neurons serving
as ``hubs'' with dense connections \cite{sporns}. (For neural connectome these
are primarily locomotor interneurons \cite{towlson}.) Thus, random mutations or
ablations would most likely hit a non-hub protein or neuron, and therefore can cause 
only a minor (i.e. local) damage to network organization.

Another manifestation of {\it C. elegans} robustness and simplicity is the fact 
that its locomotory output is low dimensional in a sense that at any given instant 
the worm's posture can be represented as a linear combination of the same four basic 
shapes, the so-called ``eigenworms'' \cite{stephens}, across different genotypes
(for wild-type and different mutants) \cite{brown}. 
In sum, all this suggests a robust control mechanism of locomotion that spans the 
three levels, form microscopic to macroscopic, and it resembles the concept of 
``robust yet fragile'' architecture that possess many engineering
and biological systems \cite{carlson}.

\vspace{0.35cm}

\noindent
{\bf Why is it difficult to integrate the knowledge of connectome and neural dynamics 
with nematode behavior?} \\
The neural connectome is a static structure over the worm adulthood. On the other 
hand, the worm behavior has a temporal aspect that can change depending on environmental 
input or internal neuronal activity. Simply saying, the same connectome can produce a 
diverse behavioral output, which means that there is no a one-to-one mapping between 
neuronal structure and behavior, or between mesoscopic and macroscopic levels 
\cite{bargmann2013}. To understand a neuronal mechanism of a particular behavior one 
has to relate it to a corresponding neural dynamics. However, there are two problems here.
One is that it is extremely difficult to record electric activity 
of {\it C. elegans} neurons because of the worm's hydrostatic skeleton that can explode
under the release of internal pressure upon dissection \cite{goodman}. This problem has 
been circumvented in recent years by the invention of optogenetic methods that enable 
imaging of calcium activity in many neurons simultaneously 
\cite{larsch,prevedel,nguyen,venkatachalam}, 
which is possible due to {\it C. elegans} transparent body (see also below). 
Calcium level in neurons is a proxy for membrane electric voltage, 
and hence there is a high hope that these type of imaging methods can yield great 
insights about neuronal control of behavior not only in the nematode but also in other
species \cite{kerr}. A second problem is related to modeling of neural dynamics. 
The activity of a neural network is to a large extent dependent on neuronal communication 
mediated by synapses and on different types of membrane ionic channels.  
Excitatory and inhibitory synapses influence network dynamics in different ways \cite{dayan},
and thus realistic modeling requires the knowledge of an identity of each synapses in
the network of interest. Unfortunately, we do not know the types of synapses in most
neurons of {\it C. elegans}, i.e., what kinds of neurotransmitters and postsynaptic
receptors are used, and whether a given synapse is excitatory, inhibitory or modulatory.
Even worse, the same neuron in {\it C. elegans} can influence differently its downstream
partners: it can excite one neuron and inhibit another \cite{chalasani,li}, and
this type of mixed signaling does not occur in more complex organisms like mammals
\cite{braitenberg}. Similarly, we know that {\it C. elegans} genome encodes a large 
repertoire of various ion channels, which can exhibit a highly nonlinear dynamics, but 
most of them remain uncharacterized \cite{wormbook}. 
Consequently, we have only a limited information on the types and properties of ionic 
channel for most neurons, and thus dynamic responses of single neurons are largely unknown
(see however \cite{goodman}). Nevertheless, we do know that {\it C. elegans} neurons, in
contrast to most vertebrate neurons, do not contain voltage-sensitive sodium channels 
(lack of genes coding for these channels \cite{bargmann1998}), and hence they do not fire 
classic Na$^{+}$-induced action potentials \cite{goodman}. Instead, these neurons can be
either depolarized in a graded fashion by Ca$^{+2}$ ions \cite{goodman,lindsay2011,mellem},
or they can fire a broad calcium action potentials ($\sim 20$ msec of width) driven by
voltage-gated calcium channels \cite{liu2018}. This indicates that {\it C. elegans} neurons
can be engaged in two different modes of communication, either analog or digital, for encoding 
external signals, which probably enhances the worm's information processing capacity. 

\vspace{0.35cm}

\noindent
{\bf In search of a biophysical model of neural dynamics and the role of optimization.} \\
Missing data for synaptic signaling and ion channel types did not, however, prevent several 
computational attempts to model neuron dynamics. One can broadly distinguish 
three classes of models for studying individual neurons: passive or simple 
\cite{karbowski2008,bryden,izquierdo2010,izquierdo2013,rakowski2013,kunert}, moderate
\cite{boyle,kuramochi}, and active or complex \cite{rakowski2017}. The passive models 
are simplified in the sense that they do not include ion channels except a linear leak 
channel. For that reason, their dynamics seems unrealistic, unless synaptic and
gap-junction inputs are so strong that they dominate over ion channel dynamics.
Indeed, there is some support for the dominant role of synaptic and gap-junction
currents at least for some locomotor interneurons, and this comes from an optimization 
of the active/complex model where calcium and calcium activated potassium channels were 
included on equal footing with the leak current \cite{rakowski2017}. This suggests
that passive models \cite{karbowski2008,bryden,izquierdo2010,izquierdo2013,rakowski2013,kunert},
although unrealistic in capturing dynamic complexity of single neurons, do not have to 
lead to wrong dynamics on a network level if synapses in this network are sufficiently 
strong. The mentioned active model \cite{rakowski2017}, involves the combination of two
dynamic equations that are highly nonlinear, one for voltage and another for intracellular
Ca$^{+2}$ concentration, in contrast to the passive models that are described only by a single
equation for the voltage. Moderate models, in turn, either consider an ion channel nonlinearities 
in an abstract way by including a switch-like term in a single voltage dynamics equation 
\cite{boyle}, or consider a spatial distribution of calcium concentration dynamics using 
three simple linear equations \cite{kuramochi}. From this, it follows that only two models,
the active one \cite{rakowski2017} and the moderate model in \cite{kuramochi}, 
can in principle be related to the dynamics of Ca$^{+2}$ in imaging experiments of freely 
moving worms \cite{larsch,prevedel,nguyen}, and thus can link computations or theory with
experimental data. Out of these two models, the active model \cite{rakowski2017} is much 
more grounded biophysically (Hodgkin-Huxley type model), and therefore has a potential 
for future improvements in adding more relevant channels and their optimization.

To summarize, we know which neuron is connected to which, but we are uncertain how 
they talk to each other and what is their intrinsic dynamic repertoire, which is a big 
obstacle for realistic modeling of worm's neural circuits and relating them to behavior. 
For these reasons, it appears that it does not make sense to model the dynamics of the 
whole large-scale {\it C. elegans} neural network, despite the knowledge of its connectome, 
because the result is very likely to be distinct from the real network activity 
(see however \cite{kunert}).
A way around these problems is to use sophisticated computational techniques relying
on evolutionary optimization that would allow us to predict or determine with high 
probability the neuronal and synaptic parameters by relating connectome to behavior
\cite{rakowski2017,izquierdo2013}. Such approaches minimize an appropriately chosen 
multivariable cost function, and find optimal set of neural network parameters for 
a given behavior. An optimization approach could be used, e.g., to infer the range of
ion channel parameters in the biophysical model used in \cite{rakowski2017}.
It seems that a natural cost function for this problem could be related to a 
difference between input-output characteristics for the experimental neuron \cite{goodman}
and theoretical neuron \cite{rakowski2017}.

\vspace{0.35cm}

\noindent
{\bf How genetic and neuronal perturbations help us to decipher neural mechanism
of worm's behavior?} \\
The advantage of possessing a well characterized nervous system and the genome is 
that we can apply perturbations to these two systems and record subsequent changes
in worm's behavior. There is a large body of evidence showing that neural connectivity
is correlated with the gene expression pattern, meaning that denser connections, 
associated with hubs, require more gene activity \cite{kaufman,fornito}. Therefore by
mutating or deleting specific genes that target specific neural connections, we
can alter the worm's behavioral output. If we additionally can quantify that output, 
we can go back and in principle infer some unknown properties of the neural circuit
underlying that particular behavior. An example of the approach along similar lines
was performed in Ref. \cite{karbowski2008}. In that work, combining genetic
perturbations with computational modeling, locomotion of mutants with altered 
synaptic transmission, channel conductance, and muscle structure was measured and 
compared with predictions of a large-scale neural circuit model. It was found, among
other things, that the head network was the likely source of worm's sinusoidal 
oscillations (the so-called Central Pattern Generator, CPG), and that the frequency 
of neuronal oscillations depends bimodally on the synaptic coupling strength, the 
latter in agreement with slo-1 mutants and their variants  \cite{karbowski2008}. 
This work is also an example that mathematical modeling does not have to act 
posteriorly to the experiment, but instead can induce experimentation by proposing 
a certain hypothesis: in this case the location of CPG. A recent experimental quest 
to resolve that question, involving genetic constructs and neural ablations, suggest 
a slightly more complex picture: there are probably more than one CPG (some perhaps 
secondary), distributed along the worm's body \cite{fouad}. Also there is an evidence
that CPG can be strongly modulated by proprioceptive feedback \cite{wen}, which can
help in coordination of various CPG's and stabilization of the worm's movement.

Cellular perturbations to a specific neural circuit can help us also in revealing 
the identity of neural connections \cite{rakowski2017,rakowski2013}. Systematic ablations 
of neurons in the pre-motor interneuron network were performed, and based on this, 
the likely polarities of interneurons and their connections were inferred using a 
computational approach, involving evolutionary optimization 
\cite{rakowski2017,rakowski2013}. Unexpectedly, it turned out that all interneuron 
connections should be inhibitory. This prediction is partly supported by recent data
showing that the connection from AVA interneuron to one of the downstream motor neurons
is mediated by lgc-46 chloride postsynaptic receptor \cite{liu2017}, which should act
as inhibition since influx of negative Cl$^{-}$ ions decreases membrane voltage
\cite{putrenko}. 
However more specific neurophysiological experiments must be performed
to verify this strong prediction. The works in Refs. \cite{rakowski2017,rakowski2013} 
have provided a substrate for a more realistic modeling of premotor interneuron network, 
and in addition they shed some light on how the locomotory output is represented on 
the interneuron level. The dominant synaptic inhibition means that interneurons mutually 
suppress one another, and this could be good for saving metabolic energy in the 
interneuron network during locomotion. Such low activity is in agreement with a study 
showing that only a small fraction of {\it C. elegans} sensory neurons is activated upon 
an external stimulation \cite{zaslaver}. Taken together, this suggests that sensory 
and premotor interneuron circuits could use sparse neural representations to encode 
locomotion \cite{rakowski2017}. Such sparseness is energetically advantageous because 
neurons and synapses are known to be metabolically expensive 
\cite{levy,laughlin,attwell,karbowski2007}.

It should be noted that a sheer perturbation like cell ablations can have ``off target''
effects \cite{otchy}. This means, e.g., that removing a certain neuron from 
a circuit of interest can affect the input coming to this circuit from other neighboring
networks, because of the possible recurrent connections between them. This undesirable
effect would be amplified if the ablated neuron were a hub neuron. It seems that it
is difficult to control this type of influence or even estimate its magnitude; nevertheless
one has to be aware of such possibilities, especially in the modeling studies.

\vspace{0.35cm}

\noindent
{\bf Optogenetics as a method for perturbing and monitoring neural dynamics.} \\
Optogenetics is an experimental method that has an equal or even more potential than 
cell ablation, since it enables a selective activation or silencing of some neurons
without affecting the network structure \cite{bernstein}. The main idea is to
express, using genetic methods, artificial ion channels or pumps in neuron's membrane 
that upon light stimulation conduct various ions (e.g. Na$^{+}$, K$^{+}$, Ca$^{2+}$,
Cl$^{-}$, H$^{+}$), leading to changes in neuronal voltages. Because the light
stimulation can be changed at will, it allows a temporal control of the circuit dynamics
\cite{bernstein}. The large-scale neural activity can be subsequently 
monitored using calcium imaging in real time 
\cite{larsch,prevedel,nguyen,venkatachalam}.

It must be stressed again, that since optogenetics affects mainly ion channels,
one has to use biophysical neuron model with realistic basic channel dynamics
in order to compare computational modeling with calcium optogenetic data.
Thus, an active neuron model, such as the one in Ref. \cite{rakowski2017}, or its
extension, is essential for a meaningful connection between theory and data.

\vspace{0.35cm}

\noindent
{\bf How locomotory program of {\it C. elegans} is encoded in neural dynamics?} \\
Large-scale optogenetic calcium imaging study has shown that worm locomotory states 
are correlated not only with the dynamics of peripheral motor and command neurons, 
but also and mostly with distributed neural dynamics across the whole nervous system 
\cite{kato}. Moreover, this global neuronal representation of locomotion is highly 
coordinated to the extent that its dynamics can be reduced to just few dimensions 
\cite{kato}. A principal component analysis associated with cross-correlations between 
neurons revealed that a given locomotion state (moving forward, backward, turning, slowing) 
can be precisely mapped into a particular region of a low dimensional neuronal state 
manifold (in the space of principal components). This manifold has a continuous and 
cyclic character, reflecting a long-term repeatable sequence of the worm motor actions 
\cite{kato}. This shows that {\it C. elegans} locomotory phases are encoded in the 
nervous system collectively in a simple, yet abstract, but a straightforward way. 
Interestingly, the low dimensionality of neural dynamics underlying locomotion is 
consistent with the low dimensionality of worm's locomotory shapes discussed earlier 
\cite{brown,stephens}. Additionally, it is very similar to neural representations
in other species \cite{churchland,cunningham}, which might suggest some form of
neuronal universality across different animals with dramatically different brain sizes.

How robust is the low dimensional neuronal state manifold? It turns out that its
topological structure is quite stable against various perturbations, including an 
inhibition of a hub locomotor interneuron (AVA), or stimulation of chemosensory neurons 
\cite{kato}. In addition, the manifold structure is also preserved during {\it C. elegans} 
starvation, which surprisingly affects only a few neurons \cite{skora}. Nevertheless, 
the global neuronal manifold can collapse to a single stationary fixed point during 
the worm's quiescence, presumably global sleep brain state, in which the majority of 
neurons throughout the brain is inactive \cite{skora}. 
Thus, it appears that the global neuronal state representing locomotion is stable
if perturbations are local, but it can be destroyed when activities of many neurons
are significantly reduced. This temporal neural organization of behavior points again 
towards the engineering concept of ``robust yet fragile'' \cite{carlson}, suggesting some 
similarities between global organization in man-made devices and evolutionary shaped 
biological systems.

Is the global neuronal manifold energy efficient? The data show that about $40 \%$
of head neurons is involved in the neuronal manifold \cite{kato,skora}, which is a
substantial figure and clearly should lead to a large metabolic expenditure on a
global neural scale. This contrasts with the above results showing sparse sensory 
representation \cite{zaslaver} and mutual synaptic suppression of premotor interneurons 
\cite{rakowski2017}, and indicates that energy might be minimized locally, but not globally. 
Moreover, the high percentage of neurons participating in the global manifold suggests 
that the worm must invest some amount of energy to perform its living functions, i.e. 
to execute its exploration-exploitation trade-off \cite{berger}, and a sheer saving of 
metabolic energy is not a viable strategy.

\vspace{0.35cm}

\noindent
{\bf Locomotory decision making in {\it C. elegans} and macroscopic level.} \\
The macroscopic level where the worm's behavior takes plays is influenced both by the
nematode nervous system and by an environmental input that can change in a stochastic
manner \cite{faumont,wen}. Interestingly, the neural network in itself can generate 
variability even to deterministic stimuli, as for instance in the olfactory circuit of 
{\it C. elegans} \cite{gordus}, and this seems to be a generic feature of neural circuits 
also in higher animals \cite{destexhe}. This implies that behavioral level of the worm 
can be described in probabilistic terms using Markov chains, with dynamically changing 
probabilities representing states with forward and backward motions and stop phases 
\cite{rakowski2017,rakowski2013,roberts}. Experimental data involving genetic mutations 
in the premotor interneurons and motor neurons indicate that transitions between locomotor 
states are driven by a difference in the activities of motor neurons controlling forward 
and backward motion \cite{kawano}. Specifically, if motor neurons belonging to class B
are more active than those belonging to class A, then the worm more likely will move
forward, and vice versa \cite{kawano}. Both motor neuron classes are modulated by the
upstream interneurons, which in turn are influenced by sensory input. Hence, the
probabilistic decision to move in a particular direction or to stop is made
collectively, and presumably with some delay, by the whole interacting large-scale neural
network, whereas the execution of the decision is performed by the asymmetric activities
of the two classes of peripheral motor neurons.

How can we reconcile this probabilistic picture of locomotor output with the above
continuous dynamics of global neuronal manifold where each locomotory state is mapped to
a specific manifold region \cite{kato,skora}?
It seems that these two descriptions are mutually complementary: 
the first is the macroscopic (behavioral) view, while the second is its abstract neuronal 
representation. It must be also remembered that the neuronal manifold representation does 
not include time explicitly, and hence stochastic temporal effects are hidden. 
Nevertheless, these effects are captured by a thickness of the neuronal manifold \cite{kato}.
Moreover, from a mathematical point of view, the probabilistic Markov chain picture can
be formally mapped into continuous dynamics of a relevant variable (e.g. velocity) perturbed
by stochastic noise \cite{vankampen}. Thus, these two, seemingly distinct, descriptions can
be made formally similar under suitable mathematical transformations (see Box 1).

Another issue is how we can use biophysical neural models to describe locomotory
transitions in a simple and efficient way, i.e., how mesoscopic level can describe
the macroscopic one? It appears that solely biophysical 
description of locomotory behavior is impractical, since there is no a unique
neural variable that relates directly to motor output. Instead, it seems that a
better approach is to use a hybrid picture: biophysical modeling of neural dynamics
combined with probabilistic Markov model for the description of behavior, as it
was done in \cite{rakowski2017}. The two models meet at the level of motor neuron
activity, which links mesoscopic and macroscopic dynamics.

\vspace{0.35cm}

\noindent
{\bf Summary and conclusions.} \\
{\it C. elegans} worms exhibit a robust low-dimensional locomotory behavior, which
can be mildly perturbed by genetic and cellular means to decipher its neuronal 
representation. However, to describe and understand the neuronal basis of locomotion, 
we need more reliable biophysical neural models, and models linking directly neural 
dynamics with different locomotor states. We also need a more specific molecular 
perturbations, which could be straightforwardly related to specific parameters in the 
models. Optogenetic methods, in combination with optimization techniques, will allow 
a comparison between predictions of neural theory with experimental data, and thus 
can select better models. Once this happens, the genuine systems biology approach, 
from genes through neurons and their networks to behavior, could be successfully applied
in {\it C. elegans}. Such combined approaches might uncover neuronal principles of
behavior in the worm that could have a universal character and apply to other, more 
complex, animals as well, since evolution acts in many respects by repeating the same 
microstructures and mechanisms \cite{kitano,hartwell}.

Finally, we did not discuss other biological functions, like learning and memory, 
which are known to exist in {\it C. elegans} \cite{ardiel,sasakura}. For instance, 
some neurons can participate in diverse behaviors, e.g. olfactory sensing and encoding, 
which is a form of memory \cite{ha,luo}. This is interesting because it links sensory 
and locomotory circuits with learning and memory, which currently lacks understanding. 
Clearly, there is a potential here for using computational modeling that would give 
us more insight about such a complex behavior.

\newpage

\noindent
{\bf \underline{Box 1: Link between probabilistic and continuous descriptions of locomotion. }} 

It is shown below that the probabilistic Markov model of {\it C. elegans} locomotion employed
in \cite{rakowski2017,roberts} is approximately equivalent to continuous noisy dynamics
of worm's velocity. The general form of the Markov model is given by dynamics of probabilities
$p_{n}(t)$ that the system is in a particular state $n$:

\begin{eqnarray}
\dot{p}_{n}= \alpha_{n+1}p_{n+1} + \beta_{n-1}p_{n-1} - (\alpha_{n}+\beta_{n})p_{n},
\end{eqnarray} \\
where the dot denotes the time derivative, and $\alpha_{n}$, $\beta_{n}$ are the transition
probabilities per time unit.
In the minimal models of {\it C. elegans} locomotion only 3 states are considered:
forward movement, backward movement, and stop phase
\cite{rakowski2013,rakowski2017,roberts}, and these states correspond respectively to
$n=1$, $n=-1$, and $n=0$ in Eq. (1). We can extend these minimal models to the model with
$2N+1$ states, where $N \gg 1$, in which $n=1,2,...,N$ correspond to the states with different
velocities for forward motion, the states $n=-1,-2,...,-N$ correspond to different velocities
for backward motion, and the state $n= 0$ refers to the stop phase.

The goal is to transform the discrete Eq. (1) into a quasi-continuous equivalent equation.
To achieve this, we define a quasi-continuous velocity $v$ as $v= n\epsilon$, where $\epsilon$
is the experimental resolution of velocity or its measurement error. The next step is to
introduce two quasi-continuous functions $A(v)$ and $B(v)$, related respectively to
deterministic and stochastic components of the dynamics, by making the following rescaling
transformations:

\begin{eqnarray}
p_{n}(t)= \epsilon P(v,t),   \nonumber \\
\alpha_{n}= \frac{1}{2}\left[ \frac{B(v)}{\epsilon^{2}} - \frac{A(v)}{\epsilon}\right],
\nonumber \\
\beta_{n}= \frac{1}{2}\left[ \frac{B(v)}{\epsilon^{2}} + \frac{A(v)}{\epsilon}\right],
\end{eqnarray}
where $P(v,t)$ is the distribution of probability for $v$. 
Note that nonzero $A(v)$ introduces asymmetry between elementary jumps that increase
($\beta_{n}$) and decrease ($\alpha_{n}$) velocity.
A quasi-continuous version of Eq. (1) is obtained by expanding $p_{n\pm 1}$, $\alpha_{n+1}$,
and $\beta_{n-1}$ for small $\epsilon$ to $O(\epsilon^{2})$ order. For example,
$p_{n\pm 1}= \epsilon P(v\pm\epsilon,t)= \epsilon[P(v) \pm \epsilon P'(v) +
  \frac{\epsilon^{2}}{2}P''(v) + O(\epsilon^{2})]$,
where the prime and double-prime denote first and second derivatives with respect to $v$.
These expansions lead to a partial differential equation for probability distribution $P(v,t)$
of quasi-continuous velocity as

\begin{eqnarray}
 \frac{\partial P(v,t)}{\partial t}= - \frac{\partial}{\partial v}[A(v)P(v,t)]
   + \frac{1}{2}\frac{\partial^{2}}{\partial v^{2}}[B(v)P(v,t)] + O(\epsilon^{2}),
\end{eqnarray} \\
which formally is known as a Fokker-Planck equation \cite{vankampen}.
This equation in turn is mathematically equivalent to a continuous dynamics of velocity $v$
influenced by noise \cite{vankampen}

\begin{eqnarray}
\dot{v}= A(v) + \sqrt{B(v)}\eta,
\end{eqnarray} \\
where $\eta$ is the Gaussian white noise with unit variance. Note that in this picture,
the function $A(v)$ is related to a deterministic part of acceleration, while $B(v)$ to its
stochastic part. Both of these functions can in principle be experimentally measured, which
should enable us to determine the transition rates $\alpha_{n}, \beta_{n}$ in the phenomenological
model given by Eq. (1).

\newpage

\vspace{2.0cm}

\noindent{\bf Acknowledgments} \\
The author thanks an anonymous reviewer for useful comments.
The work was supported by the Polish National Science Centre 
(NCN) grant no. 2015/17/B/NZ4/02600.

\newpage

\vspace{1.5cm}

$*$ of special interest

$**$ of outstanding interest

%\noindent{\bf\large  References} \\

%\vspace{1.5cm}

\newpage

{\bf \large Figure Captions}

\vspace{0.3cm}

{\bf Fig. 1} \\
{\bf Robustness of {\it C. elegans} locomotion.} 
Scaling of propulsion velocity with body wave velocity (product of undulatory 
frequency $f$ and body wavelength $\lambda$) for {\it C. elegans} across wild type 
and several mutants. Note that most of the strains align along a
common line (least square fitting; black line). Adapted and modified from 
Ref. \cite{karbowski2006}. Mutants included: cat-2(e112), cat-4(e1141), egl-30(tg26),
goa-1(n1134), goa-1(sy192), lon-1(e185), double mutant lon-1(e185);lon-2(e678),
BE109, sqt-1(sc101), sqt-1(sc103), unc-54(st130), unc-54(st132), unc-54(st134),
unc-54(st135), unc-54(s95), and unc-54(s74).

\vspace{0.3cm}

%\end{narrowtext}
\end{document}